\documentclass[prl,floatfix,twocolumn,showpacs,amsfonts,amsmath,amssymb]{revtex4}
\usepackage{graphicx} 
\usepackage{subfigure}
\usepackage{dcolumn} 
\usepackage{bm} 
\usepackage{color}
\usepackage{float}

\newcommand{\g}{\ifmmode\text{g}\else{g}\fi}

\begin{document}
	
\title{Mechanisms for strong anisotropy of
in-plane g-factors in hole based quantum point contacts}
	
\date{\today}

\author{D.~S.~Miserev$^1$, A.~Srinivasan$^1$, 
O.~A.~Tkachenko$^2$, V.~A.~Tkachenko$^{2,3}$,
I. Farrer$^{4}$, D. A. Ritchie$^5$,  A. R.~Hamilton$^1$, O.~P.~Sushkov$^1$} 
\affiliation{$^1$School of Physics, University of New South Wales, Sydney, 2033, Australia}
\affiliation{$^2$ Rzhanov Institute of Semiconductor Physics of SB RAS, Novosibirsk, 630090, Russia }
\affiliation{$^3$ Novosibirsk State University, Novosibirsk, 630090, Russia}
\affiliation{$^4$Department of Electronic and Electrical Engineering, University of Sheffield, Sheffield, S10 2TN, United Kingdom}
\affiliation{$^5$Cavendish Laboratory, University of Cambridge, Cambridge, CB3 0HE, United Kingdom}
	
\begin{abstract}
In-plane hole g-factors measured in quantum point contacts  based on p-type heterostructures
strongly depend on the orientation of the magnetic field with respect to the electric current.
This effect, first reported a decade ago and confirmed in a number of publications,
has remained an open problem.
In this work, we present systematic experimental studies to disentangle
different mechanisms contributing to the effect and develop the theory which describes it successfully.
We show that there is a new mechanism for the anisotropy related to the existence of an additional
$B_+k_-^4\sigma_+$
effective Zeeman interaction for holes, which is kinematically different from the
standard single Zeeman term $B_-k_-^2\sigma_+$ 
considered until now.

\end{abstract}
	
	\maketitle
 A quantum point contact (QPC) is a narrow quasi-one-dimensional (1D) constriction linking two two-dimensional (2D) electron or hole reservoirs.
 Experimental studies of QPCs started with the discovery of the conductance quantization in steps of $G_0=2e^2/h$~\cite{Wees1988,Wharam1988}. The steps are due to the quantization of transverse channels~\cite{Buttiker1990}. Effects of many-body correlations in QPCs were identified by a ``0.7-anomaly'' in the conductance,  an enhancement of the \g-factor  in the 1D limit~\cite{Thomas1996}, and by a zero bias anomaly~\cite{Cronenwett2002}. G-factors in n-type QPCs have been measured in numerous experiments; a relatively recent one is reported in Ref.~\cite{Burke2012}. 

The in-plane electron g-factor in a QPC takes the same value for any direction of the in-plane magnetic field. Even in InGaAs, which has appreciable spin-orbit coupling, no in-plane g-factor anisotropy has been observed~\cite{martin}. Contrary to this, measurements for holes in QPCs based on GaAs p-type heterostructures
indicate a huge anisotropy. 
All previously reported values of the g-factor for  magnetic fields applied perpendicular
to the QPC are consistent with $g_{\perp}=0$ within experimental error, while the g-factor
$g_{||}$ for  the parallel direction is nonzero~\cite{chen,komijani,nichele}.

Regardless of numerous studies, the g-factor anisotropy effect in QPCs remains unclear. One mechanism to explain the g-factor anisotropy was suggested in Ref.~\cite{komijani}. This mechanism is based on the crystal anisotropy of the cubic lattice. While it is not negligible, the contribution of this mechanism is too small to explain the observed anisotropy.

In this work, we identify a new  mechanism for the g-factor anisotropy unrelated to the crystal lattice.
It is instructive to use classification in powers of crystal anisotropy $\eta$ defined below.
The new mechanism is leading in $\eta$ and the mechanism~\cite{komijani} is subleading.
The new mechanism is negligible at very low hole densities. However, at real physical densities it is the major anisotropy mechanism.
Previous measurements were performed in 2D hole systems formed at a single heterojunction~\cite{chen,komijani}, 
which can be modeled as a triangular potential well. There is also a  measurement with an asymmetric quantum well~\cite{nichele} 
which can be modeled as a square potential with an electric field along the z-axis. 
The z-axis is perpendicular to the plane of the 2D hole  system.
The z-asymmetry results in the cubic Rashba spin orbit interaction 
(SOI)~\cite{Rashba,chesi,culcer,nichele2}. 
We will show that there are two major mechanisms for $g_{\perp}$ suppression, (i) the $g_1-g_2$-mechanism,
(ii) the Rashba mechanism.
To disentangle the mechanisms,
in the present work we perform measurements of QPC g-factors for quantum well GaAs 
heterostructures which allows us to tune the Rashba SOI.
By reducing the Rashba SOI we observe a non-zero $g_{\perp}$ for the first time (although the
anisotropy is still large, with $g_{||}\gg g_{\perp}$).
In all previous measurements the strong asymmetry of the heterostructure, or the high hole density
resulted in a very strong Rashba SOI, so both mechanisms contributed to
suppression of $g_{\perp}$.
The Rashba mechanism  is not significant in our devices. (The mechanism is
explained in the very end of the paper and discussed in detail in the supplementary material D.)
The hole gas is confined in a 15nm rectangular quantum well.
An external electric field $E_{z}$ is superimposed on the well using an in-situ back gate below the quantum well. 
\begin{figure}[h]
	\includegraphics[width=0.48\textwidth]{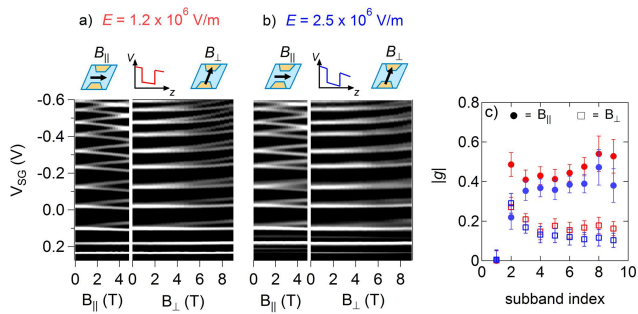}
	\caption{
Panels a,b:  Greyscale plots of the transconductance showing Zeeman spin splitting of 1D hole subbands 
in a magnetic field applied parallel and  perpendicular the QPC.
Panel a (Panel b) corresponds to the electric field along the z-axis $E_z= 1.2 \times 10^6$V/m 
($E_z= 2.5 \times 10^6$V/m).
Panel c: Absolute values of  the subband g-factors extracted from data in Panels a and  b.
The circles (squares) correspond to the direction of
    magnetic field along (perpendicular)  to the QPC.
The red (blue) symbols correspond to the out-of-plane electric field  $E_z= 1.2 \times 10^6$V/m 
($E_z= 2.5 \times 10^6$V/m).
	}
	\label{QPC}
\end{figure}
The transconductance maps measured at $E_z= 1.2 \times 10^6$V/m and $E_z=2.5 \times 10^6$V/m 
are presented in Fig.\ref{QPC}a,b. 
The absolute values of the g-factors extracted from these maps are shown in Fig.\ref{QPC}c.
All experimental details are provided in Section A of the supplementary material, see also Refs.~\cite{ClarkeJAP06,Nextnano,PatelPRB91}.
Fig.\ref{QPC} demonstrates  a significant g-factor anisotropy.
Another observation is that in all cases both g-factors are very small for the lowest transverse channel, 
$n = 1$.

Dynamics of a single hole in bulk conventional semiconductors are described 
by the Luttinger Hamiltonian~\cite{luttinger}. We consider here the spherical approximation~\cite{baldereshi}
\begin{eqnarray}
H_L = \left(\gamma_1 + \frac{5}{2} {\overline{\gamma}}_2 \right)
\frac{{\bf p}^2}{2 m} 
- \frac{{\overline{\gamma}}_2}{m} \left({\bf p} \cdot {\bf S} \right)^2 \ ,
 \label{Lut1}
\end{eqnarray}
where ${\bf p}$ is the 3D quasi-momentum; ${\bf S}$ is the spin $S = 3/2$; 
$\gamma_1$, ${\overline{\gamma}}_2=(2\gamma_2+3\gamma_3)/5$
are Luttinger parameters; $m$ is the free electron mass.
There is also a non-spherical  part of the  Luttinger Hamiltonian
that depends on the cubic lattice orientation. 
This part is proportional to $\eta=(\gamma_3-\gamma_2)/{\overline{\gamma}}_2$.
 The parameter $\eta$ is small in compounds with
large SOI, for example  $\eta=0.34$ in GaAs and $\eta=0.09$ in InAs.
In Ref.~\cite{komijani} a mechanism of the QPC g-factor asymmetry due to 
the non-spherical part of the Luttinger Hamiltonian was suggested.
 The contribution of this mechanism is small and
is calculated in Section B of the supplementary material, see also Refs.~\cite{Miserev1,simion,Marie}.
Here we concentrate on the leading contribution which arises  from the  spherical Hamiltonian (\ref{Lut1}).

A quantum well potential $W(z)$ imposed on (\ref{Lut1}) 
confines dynamics along the z-axis leading to 2D subbands.
Here, we consider only the lowest sub-band with dispersion
\begin{equation}
\label{h0}
H_0=\varepsilon_{\bm k} \ ,
\end{equation}
where ${\bm k}=(k_x,k_y)=(p_x,p_y)$ is the 2D momentum. 
At $k=0$ the projection of spin on the z-axis $S_z$ is a good quantum number.
Due to the negative sign of the second term in (\ref{Lut1}), the lowest band
is a Kramers doublet with $S_z = \pm 3/2$.
The standard way to describe
the Kramers doublet is to introduce the effective spin $s=1/2$ with
related Pauli matrices ${\bm \sigma}$. The correspondence at $k=0$
is very simple: $|\uparrow\rangle=|S_z=3/2\rangle$,
$|\downarrow\rangle=|S_z=-3/2\rangle$.
Note that the effective spin operators  $\sigma_\pm = \sigma_x \pm i \sigma_y$ flip
$S_z = \pm 3/2$ projections. Hence, $\sigma_\pm$ are transformed  as $S_\pm^3$.

Now we apply in-plane magnetic field ${\bf B}$.
The kinematic structure of the effective 2D Zeeman Hamiltonian is of the form~\cite{Miserev1}
\begin{eqnarray}
\label{heff}
H_Z=&-&\frac{\mu_B}{4}\left\{{\overline g_{1}}[B_+k_+^2\sigma_-+B_-k_-^2\sigma_+]\right.\nonumber\\
&+&\left.{\overline  g_{2}}[B_-k_+^4\sigma_-+B_+k_-^4\sigma_+]\right\}  \nonumber\\
g_{1}(k) &=& k^2  {\overline g_1}(k) \ , \ \ \   g_{2}(k) = k^4  {\overline g_2}(k) \ .
\end{eqnarray}
Pauli matrices $\sigma_\pm$   ($\sigma_\pm^2=0$)
have the angular momentum selection rule $\Delta J_z=\pm 3$, and $B_\pm$ corresponds to 
$\Delta J_z=\pm 1$. The powers of $k_{\pm}$ in (\ref{heff}) balance  the z-component of the angular 
momentum in such a way that the total Hamiltonian conserves the angular momentum,   $\Delta J_z=0$.
While the $g_1$-term in (\ref{heff}) is well known, the $g_2$-term has never been considered before.
In perturbative treatment of the Luttinger Hamiltonian (\ref{Lut1}) the $g_2$-term
appears only in a  high order of the perturbation theory.
Of course, at small momenta  $g_2 \ll g_1$, practically this is true if $kd < 0.6$, where
d is the width of the well. However, all experiments we are aware of (including ours) are
performed at $kd > 1.2$. In this case $g_1$ and $g_2$ are comparable.

\begin{figure}[h]
	\includegraphics[width=0.48\textwidth]{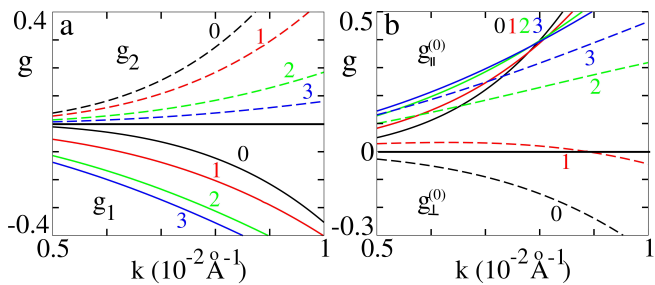}
	\caption{Panel a: Functions $g_1$ (solid lines) and $g_2$ (dashed lines) for the 2D system 
versus momentum,  see Eqs.(\ref{heff},\ref{truegauge}).
Panel b: Functions $g_{||}^{(0)}$ (solid lines) and $g_{\perp}^{(0)}$ (dashed lines) for the 1D system versus momentum, see Eqs.(\ref{ZZ}).
Both panels  are calculated for rectangular GaAs quantum well of width $d = 15$ nm with a superimposed 
electric field $E_z$. We present plots for $E_z=0,1,2,3$ MV/m, with
black, red, green, and blue lines respectively.
The value of the electric field in MV/m
is pointed out near each line.
	}
	\label{nu}
\end{figure}

The functions $g_1(k)$ and $g_2(k)$ have been calculated recently for symmetric 
heterostructures~\cite{Miserev1}. Here we calculate them for asymmetric ones.
These functions for an infinite rectangular GaAs quantum well of width $d = 15$ nm with superimposed electric field $E_z$
are plotted in Fig.\ref{nu}a. 

Remarkably the existence of two isotropic g-functions leads to an anisotropy of the QPC g-factor.
The QPC g-factor is determined experimentally by the splitting of the transconductance peaks in a magnetic
field, see Fig.\ref{QPC}. 
We define the x-axis to be along the QPC (the direction of the current) and the y-axis perpendicular to the QPC.
The transconductance peaks correspond to the chemical potential aligning with the 1D subband edges, 
where $k_x=0$. 
Therefore, in the g-factor measurements
$k=k_y$ and $k_\pm=\pm ik$. Hence, at the 1D subband edge the Zeeman interaction (\ref{heff}) for a QPC reads
\begin{eqnarray}
\label{ZZ}
&&H_Z\to-\frac{\mu_B}{2}\left\{g_{||}^{(0)}B_x\sigma_x+g_{\perp}^{(0)}B_y\sigma_y\right\}\\
&&g_{||}^{(0)}(k)=g_2(k)-g_1(k)\ , \ \ \ g_{\perp}^{(0)}(k)=-g_2(k)-g_1(k) \ . \nonumber
\end{eqnarray}
The superscript $(0)$ indicates that these are terms of the zero order in $\eta$.
Plots of $g_{||}^{(0)}(k)$ and $g_{\perp}^{(0)}(k)$ for an infinite 15nm rectangular GaAs quantum well with different values of $E_z$ are presented in 
Fig.\ref{nu}b.

Calculations of the in-plane Zeeman response have  a nontrivial pitfall related to gauge invariance.
This pitfall was overlooked in previous studies.
To find the g-functions we diagonalize the 3D Hamiltonian 
 \begin{eqnarray}
\label{A}
H&=&H_L+W(z) - 2 \kappa \mu_B \, {\bm B} \cdot {\bm S}\nonumber\\
{\bm A}&=&[B_y(z-z_0),-B_x(z-z_0),0]\ ,
\end{eqnarray}
where ${\bm A}$ is the vector potential included in $H_L$ via ``long derivatives'' (for details see Ref.~\cite{Miserev1}), and $2\kappa$ is the bulk g-factor.
In Eq.(\ref{A}) $z_0$ is an arbitrary constant. Due to gauge invariance, $z_0$ cannot affect any physical 
observable.
However, at arbitrary $z_0$  the minimum of the 2D hole dispersion is generally not at $k=0$.
In particular, in this situation the
transconductance peaks do not correspond to $k_x=0$. To avoid this complication we fix the value of $z_0$
with the condition that the minimum of the dispersion is at $k_x=0$.
For a symmetric quantum well $W(z) = W(-z)$ the value of $z_0$ is dictated by symmetry, $z_0 = 0$, the center of symmetry of the well.
In the next paragraph we discuss how to determine $z_0$ for an  asymmetric heterostructure, $W(z)\ne W(-z)$.

An asymmetric quantum well gives rise to Rashba SOI
\begin{equation}
\label{hR}
H_{R}=-\frac{i}{2}\alpha_k (k_+^3 \sigma_- - k_-^3 \sigma_+)\ .
\end{equation}
This term has to be added to the effective 2D Hamiltonian $H_{2D}$ given by Eqs.(\ref{h0}),(\ref{heff}).
Besides the Rashba SOI (\ref{hR}) one more kinematic structure in the
effective 2D Hamiltonian is possible
\begin{equation}
\label{hB}
H_B({\bf k})=\gamma_k([{\bf B}\times {\bf k}]\cdot {\bf \hat{z}})\ .
\end{equation}
Here, $\gamma_k$ is a momentum dependent coefficient.
To the best of our knowledge, the term (\ref{hB}) was unknown in previous literature. 
The momentum independent part of $\gamma_k$ can be gauged out, see below,
hence $\gamma_k\propto k^2$  and $H_B$ scales as $k^3$ similar to (\ref{hR}). 
According to our calculations, (\ref{hR}) and (\ref{hB}) become comparable at $B\approx 10$T.
Note that (\ref{hR}) and (\ref{hB}) are the only inversion asymmetric kinematic structures allowed by 
other symmetries in the effective 2D Hamiltonian in the spherical ($\gamma_3=\gamma_2$) approximation.
The  term (\ref{hB}) can be absorbed in the dispersion, 
$\varepsilon_{\bf k}+H_B({\bf k})\approx \varepsilon_{{\bf k}+{\bf q}}$, where ${\bf q} = -m^*(k) \gamma_k [{\bf B}\times {\bf \hat{z}}]$ and $m^*(k)=k/\left(\frac{\partial \varepsilon_k}{\partial k}\right)$ is the effective mass.
This shift is equivalent to the variation of $z_0$ discussed in the previous paragraph.
To fix the dispersion minimum at $k=0$ one needs to set $\gamma_{k=0}=0$.
The value of $z_0$ providing this condition follows from the equation
\begin{equation}
\left< \left(\frac{\partial H}{\partial {\bm k}}\right)_{{\bm k}=0} \right>= \left.\frac{\partial H_{2D}}{\partial {\bm k}}\right|_{{\bm k}=0} = 0 \ .
\label{truegauge}
\end{equation}
Here $H$ is given by Eq.(\ref{A}) and $H_{2D}$ is the effective 2D Hamiltonian
which includes terms (\ref{h0}),(\ref{heff}),(\ref{hR}) and (\ref{hB}).
Brackets stand for the averaging over the wave function corresponding to
$k=0$, but $B\ne 0$. Solving Eq.(\ref{truegauge}) in the linear in B approximation
yields the value of $z_0$.
The effect of quantum well asymmetry on the 2D functions $g_1(k)$, $g_2(k)$,  and the 1D g-factors 
$g_{\parallel}^{(0)}(k)$ and $g_{\perp}^{(0)}(k)$ 
calculated with the constraint (\ref{truegauge})
for electric fields $E_z=1,2,3$ MV/m are  shown in Fig.~\ref{nu} by the coloured lines. The corresponding values of $z_0$ determined from Eq.(\ref{truegauge}) are $z_0(nm)=1.38, 2.37, 3.03$
(zero is in the center of the square well). 

To complete the discussion of gauge invariance, we would like to demonstrate that in the  presence
of the Rashba interaction (\ref{hR}) the functions $g_1$ and $g_2$ in Eq.(\ref{heff}) are not gauge
invariant. Let us perform  the shift gauge transformation
${\bm k} \to {\bm k} - \delta{\bm A}_0$, $\delta{\bm A}_0 = -\delta z_0[{\bm B}\times{\bf \hat{z}}]$.
Hence the dispersion (\ref{h0}) is changed as 
$\varepsilon_k \to \varepsilon_{{\bm k}-{\delta{\bm A}_0}}\approx \varepsilon_k-
\frac{\partial \varepsilon_k}{\partial {\bm k}}\delta{\bm A}_0$.
The $\delta {\bm A}_0$ term in this equation can be transferred to Eq.(\ref{hB}) leading 
to a change of $\gamma_k \to \gamma_k - \delta z_0/m^*$ that is discussed in the previous paragraph. One must also perform the shift  of
${\bm k} \to {\bm k} - \delta {\bm A}_0$ in the Rashba interaction (\ref{hR}),
$H_R({\bm k})\to H_R({\bm k}-\delta{\bm A}_0) \approx H_R({\bm k})-\frac{\partial H_R}{\partial {\bm k}}
\delta{\bm A}_0$.
The $\delta {\bm A}_0$ term in this equation can be transferred to Eq.(\ref{heff}), leading 
to the change $\mu_B \bar{g}_1 \to \mu_B \bar{g}_1 + \delta z_0 (6 \alpha_k + k \alpha'_k)$, 
$\mu_B \bar{g}_2 \to \mu_B \bar{g}_2 - \delta z_0 \alpha_k'/k$. 
Here $\alpha'_k=\frac{\partial \alpha_k}{\partial k}$ is the derivative of the Rashba coupling coefficient. 
Thus, the functions $g_1$ and $g_2$ are not gauge invariant.
Of course, physical g-factors are gauge invariant, but generally they are different
from $g_1$, $g_2$. Only in the gauge fixed by Eq.(\ref{truegauge})
the physical g-factors do coincide with $g_1$, $g_2$.
The same is true for the subleading corrections ${\overline {\delta}_1}$ and ${\overline {\delta}_2}$
proposed in \cite{komijani} and calculated in Section B of the supplementary material.

\begin{figure}[h]
	\includegraphics[width=0.48\textwidth]{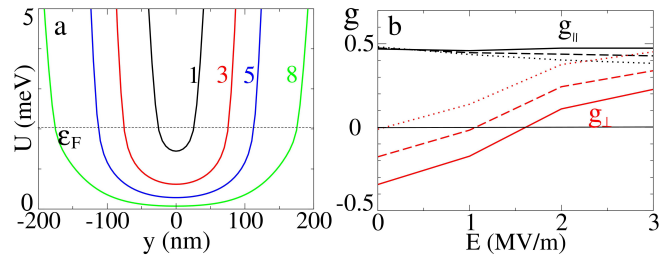}
	\caption{a): Self-consistent QPC transverse potential for 1D channels with
$n = 1, 3, 5, 8$ subbands occupied.
b): QPC g-factors $g_{\parallel}$ and $g_{\perp}$ for $n \geq 3$ versus electric field.
The heterostructure is  modeled as a 15 nm wide infinite rectangular quantum well with a superimposed electric field $E_z$.
The hole density in the 2D leads is $1.1 \times 10^{11}$ cm$^{-2}$.
The dashed lines account only for the leading spherical contribution.
The solid ([110] QPC orientation) and dotted ([100] QPC orientation) lines account for the leading 
contribution and for the first subleading one proportional to
$\gamma_3-\gamma_2$.
	}
	\label{snu}
\end{figure}

Our experiments have been performed with a 
2D hole density of $1.1 \times 10^{11}$cm$^{-2}$. It corresponds to 
a  2D Fermi momentum $k_F^{2D}=0.83 \times 10^{-2}$\AA$^{-1}$. 
The QPC channel is defined by the ``transverse" Hamiltonian,
$H_{tr}= \varepsilon_k +U(y)$ ,  $k=k_y$,
where $U(y)$ is the transverse self-consistent potential of the QPC.
The energy levels of this Hamiltonian
$E_n$, enumerated by index $n=1,2,3,...$, correspond to 
the 1D transverse channels.
Varying the split-gate voltage adjusts the self-consistent potential $U(y)$,
providing the condition to depopulate the $n^{th}$ 1D subband,
$E_n=\varepsilon_F$ .
This implies that $U(y)$ depends on {\it n}.
The self-consistent potential $U(x,y)$ for our device is calculated in Section C of 
the supplementary  material using the Thomas-Fermi-Poisson method, see Refs.~\cite{pyshkin,davies,modeling}. 
The potentials $U(y)=U(x=0,y)$ for $n=1,3,5,8$ are plotted in Fig.\ref{snu}a.

While for $n \geq 3 $ the potential minimum in the 1D channel is practically zero,
 $U(0)\approx 0$,
for $n=1$ the value of $U(0)$ is large, just slightly smaller than the Fermi energy.
Therefore, $k_y$ in this case is much
smaller than the Fermi momentum in the 2D reservoirs.
Since the in-plane g-factors scale roughly as $k_y^2$, the large value of $U(0)$
explains the very small values of g-factors for $n=1$, see Fig.~\ref{QPC}c.
Note that the potentials in  Fig.\ref{snu}a 
are very close to those obtained a long time ago
for electrons~\cite{Laux}. 
Note also that the behavior of g-factors at $n=2$ is different from that at  $n \geq 3$ and from $n = 1$,
see Fig.\ref{QPC}c.
This is because of two competing and comparable effects, (i) the reduction of g-factors since  $U(0) > 0$,
(ii) the enhancement of g-factors  due to many body Coulomb interaction effects.
The low {\it n} enhancement of the in-plane g-factor due to many body effects is well known in
 electron systems~\cite{Thomas1996}.
Fortunately, the both complications become irrelevant at $n \geq 3$.
The condition  $U(0)\approx 0$ holds, and the  Coulomb interaction is sufficiently screened.
The g-factors at $n \geq 3$
can be determined from Fig.~\ref{nu}b by taking the values  at $k=k_y= k_F^{2D}$.
This gives the g-factors $g_{||}^{(0)}$ and  $g_{\perp}^{(0)}$  shown by the 
 dashed lines in Fig.~\ref{snu}b, plotted versus the applied electric field.

To complete the story
we have also taken into account the subleading $\eta$-correction due to crystal anisotropy
proposed in Ref.~\cite{komijani}. 
We have corrected the calculations of Ref.~\cite{komijani} for some errors 
as described in the supplementary material. 
The $\eta$-correction can be described  by two momentum dependent
functions $\delta_+(k)$ and $\delta_-(k)$ defined 
in Section B of the supplementary material.
The $\eta$-correction depends on the orientation of the QPC with respect to the crystal axes
as given by Eq. (B7). In our experiment the QPC is oriented along the (110) direction,
hence the angle $\phi$ defined by Eq.(B5) is $\phi=\pi/4$. Therefore, according to (B7)
$g_{||}=g_{||}^{(0)}-\delta_-$ and $g_{\perp}=g_{\perp}^{(0)}+\delta_+$.
The plots of $\delta_{\pm}$ versus electric field are presented in panel B of Fig.B1
in the supplementary material.
Hence we arrive at the plots of $g_{||}$ and $g_{\perp}$ versus electric field shown
in Fig.\ref{snu}b by the solid black and red lines. The calculated value of $g_{\parallel}$ is practically independent of the field, and is equal to $g_{\parallel}\approx 0.46$.
 In contrast, the perpendicular g-factor $g_{\perp}$ depends on the field significantly, and even changes 
sign. However, at values of the field used in the experiment, $E_z=1.2$ MeV/m and $E_z=2.5$ MeV/m
the absolute values of the g-factor are practically equal, $|g_{\perp}|\approx 0.17$.
The theory agrees with data presented in Fig.\ref{QPC}.
We stress that in $g_{\parallel}$ there is no compensation between different
contributions, therefore the calculation is rather reliable.
On the other hand, for $g_{\perp}$ there is a significant compensation between 
the $g_1$- and $g_2$-contributions,
therefore the expected theoretical uncertainty in $g_{\perp}$ is larger than that
in $g_{\parallel}$.
Dotted lines in Fig.\ref{snu}b show our prediction for the [100] orientation of the QPC.
The essential ingredients of the theory are the functions $g_1(k)$, $g_2(k)$ considered in the
main text and the coefficients $\delta_{\pm}$ calculated in the supplementary material.
In principle, one can disentangle these parameters experimentally  by performing measurements for different $E_z$ with
a set of QPCs aligned along different crystal orientations. 
Ideally the electric fields should encompass the values shown in Fig.\ref{snu}b, with QPC's oriented
along the [110] and [100] directions.
All the devices must have the same density of holes in leads.

Besides the $g_1-g_2$ effect considered above,  and the crystal anisotropy $\eta$-correction
calculated in Section B
of the supplementary material, there is one 
more effect influencing $g_{\perp}$. 
This 1D effect is due to a combination of the transverse QPC confinement with the
Rashba SOI. It was previously addressed in numerical calculations for hole~\cite{Gelabert} and electron~\cite{kolasinski} wires. 
The 1D effect leads to $g_\perp$ oscillations and suppression with subband number $n$, $\propto \frac{\sin \pi n \delta_R}{\pi n \delta_R}$,
where $\delta_R$ is a parameter related to the Rashba SOI. At the same time, $g_\parallel$ is not affected.
This effect is weak in quantum wells, and hence is irrelevant for our experiments,
but is relevant in other experiments~\cite{chen,komijani,nichele}.
 The effect is discussed in Section D of the supplementary material.

{\it In conclusion} We have performed systematic experimental and theoretical studies to resolve the
problem of anisotropic g-factors measured in quantum point contacts based on p-type heterostructures.
We found that the most important mechanism for the anisotropy is related  to the existence of 
two kinematically different   effective Zeeman interactions for holes.  
 Using our theory we make several predictions to motivate further experiments. 
The predictions include the effects of: 
(i) Variation of density in  the leads (Fig. 2b), (ii) Change of the QPC orientation (Fig. 3b),  and
(iii) Variation of the electric field $E_z$ (Fig 3b).

We thank Tommy Li and Stefano Chesi for important discussions.
The device used in this work was fabricated in part using facilities of the NSW Node of the Australian National Fabrication Facility.
The work has been supported by the Australian Research Council grants DP160100077 and DP160103630.

 \end{document}


\renewcommand{\theequation}{A\arabic{equation}}
\renewcommand{\thefigure}{A\arabic{figure}}
\setcounter{equation}{0}
\setcounter{figure}{0}

\preprint{supplemental material}

\title{Supplemental material: Mechanisms for strong anisotropy of in-plane g-factors in hole based quantum point contacts}

\author{D.~S.~Miserev$^1$, A.~Srinivasan$^1$, 
O.~A.~Tkachenko$^2$, V.~A.~Tkachenko$^{2,3}$,
I. Farrer$^{4}$, D. A. Ritchie$^5$,  A. R.~Hamilton$^1$, O.~P.~Sushkov$^1$} 
\affiliation{$^1$School of Physics, University of New South Wales, Sydney, Australia}
\affiliation{$^2$ Rzhanov Institute of Semiconductor Physics of SB RAS, Novosibirsk, 630090, Russia }
\affiliation{$^3$ Novosibirsk State University, Novosibirsk, 630090, Russia}
\affiliation{$^4$Department of Electronic and Electrical Engineering, University of Sheffield, United Kingdom}
\affiliation{$^5$Cavendish Laboratory, University of Cambridge, United Kingdom}

\date{\today}

\maketitle

\section{A: measurement}
\subsection{Device}

The device used in this work is fabricated from a GaAs/AlGaAs accumulation mode heterostructure \cite{ClarkeJAP06} grown on a 
(100)-oriented GaAs substrate (Wafer W713).   The 2D hole system (2DHS) is formed in a 15nm GaAs quantum well (QW), 85nm below 
the surface.  The heterostructure features a n+ Si doped GaAs layer 1.2$\mu m$ below the 2DHG, acting as an in-situ backgate. 
A topgate is also patterned on the surface allowing the electric field across the quantum well (and hence the Rashba SOI) to be 
tuned at constant hole density (see device schematic in Fig.A1a).  

In our experiment, the device was measured at two gate voltage settings corresponding to the maximum and minimum attainable 
electric fields (limited by gate leakage) at a 2D hole density $n \sim 1.1 \times 10^{11} cm^{-2}$.
Below we call these setup S1 and setup S2: 
	\begin{tabular}{| C{0.9cm}| C{1.8cm} | C{1.6cm} | C{1.6cm} | C{1.8cm} |}
	\hline
	setup&\textit{E} ($10^6$~V/m) & topgate (V) & backgate (V) & density ($10^{11} cm^{-2}$) \\  
	\hline
	S2& $2.5$ & -1.43 & +0.9 & $1.1$ \\  
	S1& $1.2$ & -1.20 & -0.9 & $1.1$ \\ 
	\hline
	\end{tabular}\\
To quantify the electric field along the z-axis in our experiment, the QW potential was modelled using a self consistent 
Poisson-Schr{\"o}dinger solver (Nextnano++ software \cite{Nextnano}).  Fig.A1b shows the QW potential and 
heavy-hole  wavefunction probabilities for the two gate voltage settings used in 
the experiment.  
The electric field $E$ was estimated as average of the electric fields at the
 walls of the QW.
This gives: $E = 1.2 \times 10^6~V/m$ and $2.5 \times 10^6~V/m$ for the two gate voltage settings used in the experiment.  
\begin{figure}\
\includegraphics[width=0.99\linewidth]{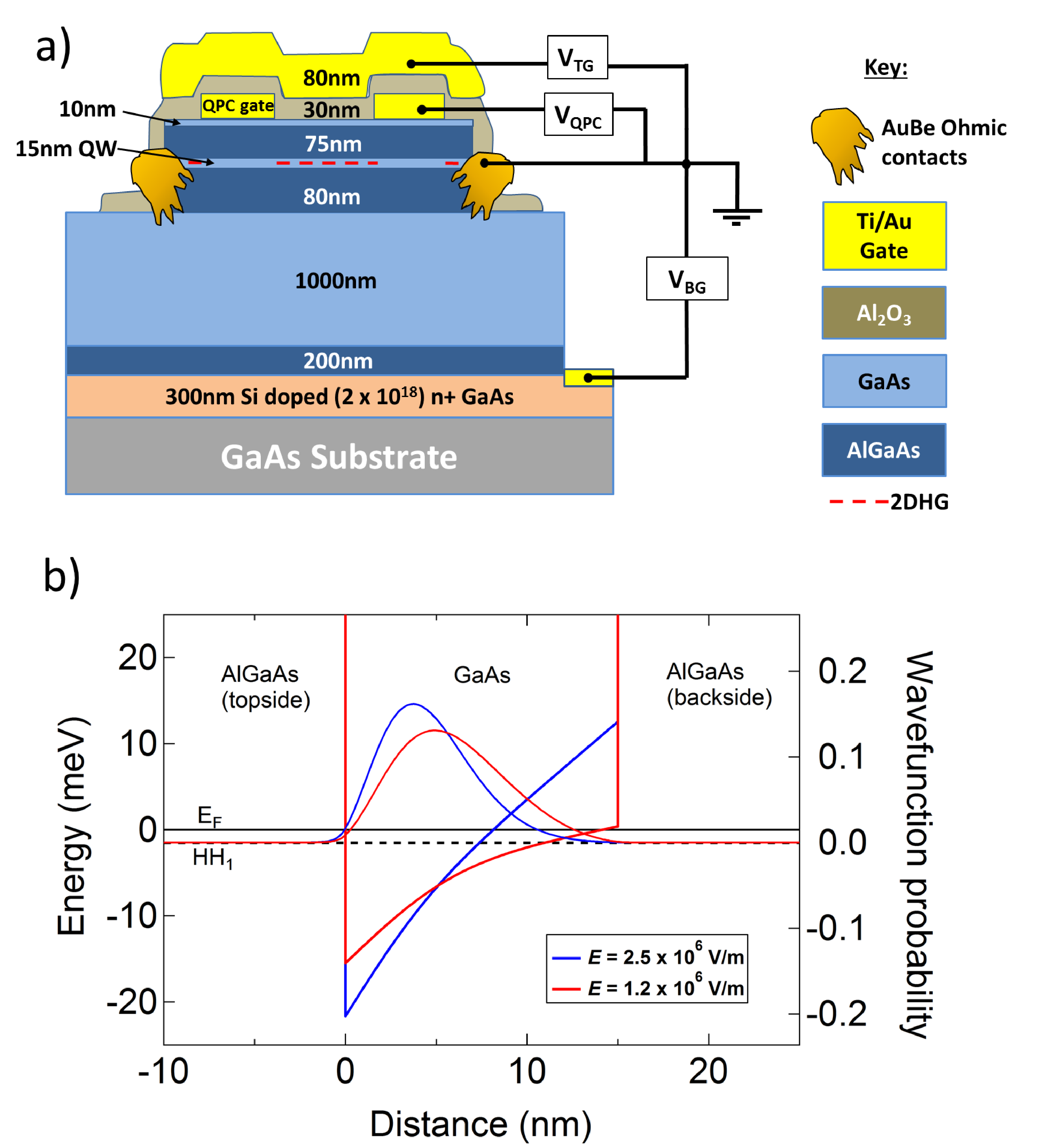}
\caption{a) Schematic diagram of device showing how gate biases are applied to control the electric field across the QW. b) QW confinement potential (bold) and heavy-hole wavefunction probabilities plotted for two gate voltage settings. The bold black line shows the Fermi energy, and the dotted black line shows the energy of the first heavy-hole band.  
The electric field   was estimated as average of electric fields at walls of the 
quantum well,
 $E_z = 1.2 \times 10^6 V/m$ and $E_z = 2.5 \times 10^6 V/m$.
} 
\label{FSU}
\end{figure}

The 2DHS is further confined using split-gates to form a 400nm wide $\times$ 300nm long QPC, with current along the 
[$01\overline{1}$] crystal direction.
The actual width of the QPC conducting channel varies from 50 to 350nm depending on the subband, see Fig.3A in the main text.
The hole density under the split-gates is zero, hence the configuration of electric field $E_z$
varies within the QPC.
The QPC g-factors depend on the value of the electric field $E_z$ in the quantum well
at the QPC neck.
On the other hand the values of the field presented in the Table above and in 
Fig.\ref{FSU}b
have been calculated for an infinite 2D system.
This implies that they are valid only far away from the QPC.
In Section C of the supplementary material we solve the full 3D electrostatic problem
using the Thomas-Fermi-Poisson method and show that the field at the neck is
practically equal to that far away from the QPC.

\begin{figure}
\includegraphics[width=0.99\linewidth]{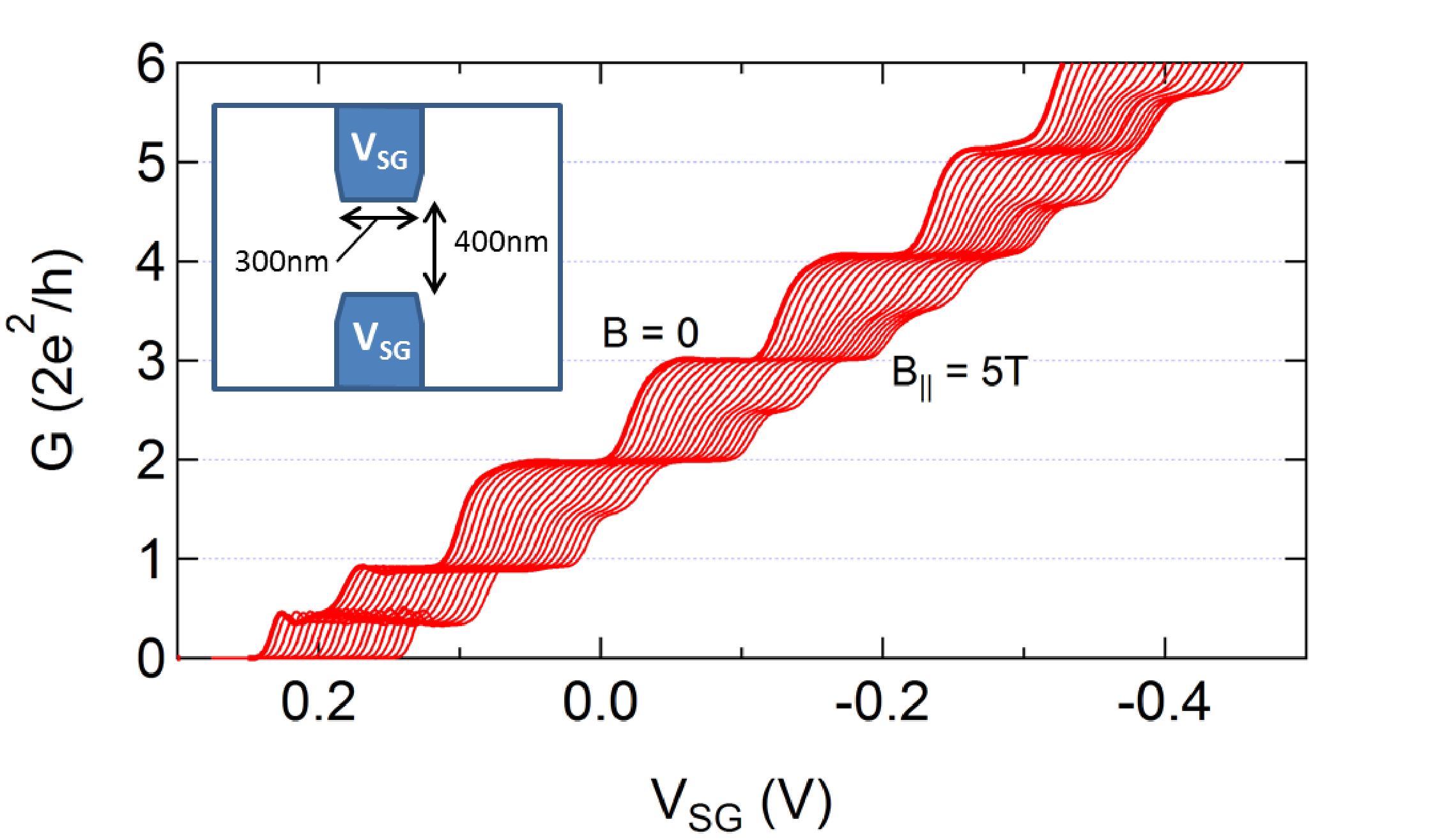}
\caption{Conductance as a function of side gate voltage showing characteristic 1D plateaus at B = 0 (bold), which evolve in magnetic field to form spin resolved half plateaus.  The traces are offset for clarity.}
\label{S2}
\end{figure}

Measurements were carried out in a dilution refrigerator, with a base temperature below 40mK, using standard ac lock-in techniques with a $100\mu V$ excitation at 77Hz.  A three-axis vector magnet was used to accurately control the direction of the in-plane magnetic field with respect to the QPC.  Fig.\ref{S2} shows the conductance as the QPC is pinched off, revealing clean 1D conductance plateaus in units of $2e^2/h$ at B = 0, which evolve to spin resolved half plateaus when a magnetic field is applied parallel to the QPC.  

\subsection{Zeeman splitting of 1D hole subbands}

Zeeman splitting of the 1D hole subbands was measured with an in-plane magnetic field applied parallel and then 
perpendicular to the QPC, corresponding to measurements of $\vert g_{\parallel}\vert$ and $\vert g_{\perp}\vert$.  
This was carried out for two settings of the electric field, $E_z$: $1.2 \times 10^6 V/m$ and 
$2.5 \times 10^6 V/m$. 
The experimental data is presented in Fig.1a in the main text
as grey-scale plots of the trans-conductance ($\partial G/\partial V_{SG}$), where light regions represent high 
transconductance, corresponding to the 1D subband edges.
The Zeeman spin splitting of each 1D subband as a function of magnetic field is clearly visible. 
It is evident that the measured Zeeman splitting is larger with the magnetic field applied parallel to the QPC, 
compared to perpendicular, suggesting $\vert g_{\parallel}\vert > \vert g_{\perp}\vert$.  
The data is summarized in Fig.1 in the main text.

\subsection{g-factor calculation using DC source drain bias spectroscopy}

The $g$-factor was extracted by measuring the Zeeman splitting in gate voltage $\Delta V_{SG}(B)$, which is then converted to a Zeeman energy splitting $\Delta E(B)$ using the well known DC source drain bias spectroscopy technique \cite{PatelPRB91}:
\begin{equation}
\Delta E(B)~=~\Delta V_{SG}(B) \times \frac{e\partial V_{SD}}{\partial V_{SG}}\\
\end{equation}
Here $\Delta V_{SG}(B)$ is the splitting of the subbands in gate voltage in a magnetic field, and $(\partial V_{SD}/ \partial V_{SG})$ is the so called “lever arm”, which converts the splitting in gate voltage to a splitting in energy.  To obtain the lever arm, we apply a dc bias between the source and drain contacts and measure the conductance at B = 0.  When a dc bias ($V_{SD}$) is applied, the transconductance peaks split in gate voltage by an amount proportional to $eV_{SD}$, giving the lever-arm $(\partial V_{SD}/ \partial V_{SG})$.

\begin{figure}
\includegraphics[width=0.99\linewidth]{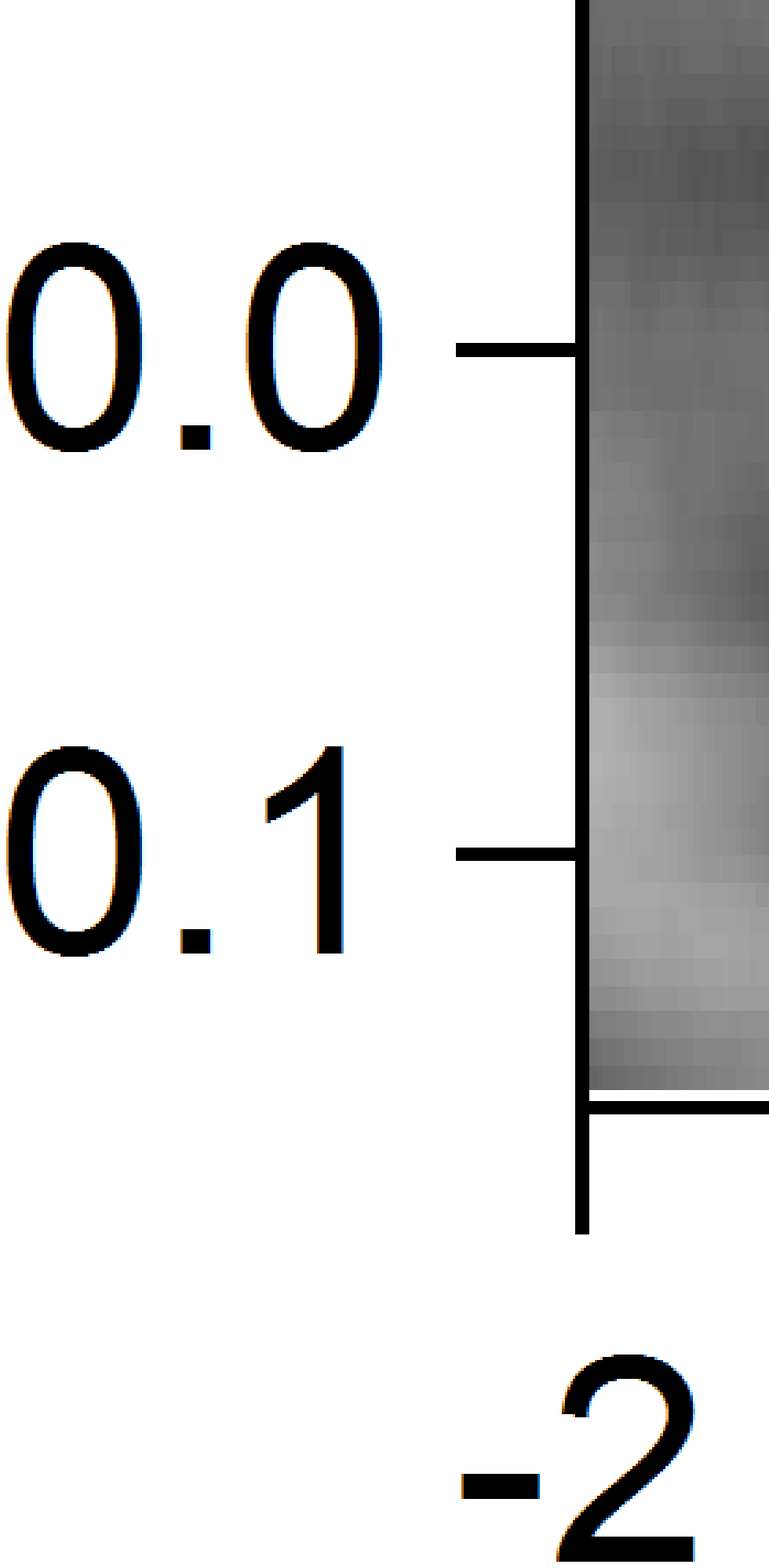}
\caption{a) Greyscale plots of the transconductance, showing 1D subband edges splitting as a function of DC source drain bias.  b) lever arm extracted from a).} 
\label{sdf}
\end{figure}

A colour plot of the transconductance $(\partial G/ \partial V_{SG})$ as a function of the gate voltage (y axis) and DC source drain bias (x axis) is presented in Fig.\ref{sdf}a. The lever arm is corrected for the series resistance, and the red traces show the actual dc bias dropped across the 1D channel when the applied bias is 3mV.  The lever-arm was calculated for each subband (each of which is at a different side gate voltage), as shown in Fig.\ref{sdf}b.

The measured Zeeman splitting in gate voltage $\Delta V_{SG}(B)$ from Fig.1a,b in the main text
 is combined with the lever arms extracted from Fig.\ref{sdf}b to obtain the g-factors given 
in Fig.1c in the main text.

\renewcommand{\theequation}{B\arabic{equation}}
\renewcommand{\thefigure}{B\arabic{figure}}
\setcounter{equation}{0}
\setcounter{figure}{0}

\section{B: In-plane Zeeman interaction proportional to the first power of  $\eta$}
In GaAs the value of $\eta = (\gamma_3 -\gamma_2)/\overline{\gamma}_2$ is $\eta=0.34$.
In the 3D Luttinger Hamiltonian terms linear in $\eta$ arise from a $4^{\textrm th}$ rank tensor. If the z-axis is directed along a crystal axis, then only $\pm 4$ projections of the tensor contribute to the term which is not invariant over rotations around the z-axis (see Eq. (59) in Ref.~\cite{simion}):
$$
H_{ten} = \frac{\eta \overline{\gamma}_2}{8 m} \left( k_+^2 S_+^2 + k_-^2 S_-^2 \right).
$$
Due to this the corresponding effective 2D Zeeman Hamiltonian, bilinear in spin and in-plane magnetic field and linear in $\eta$, must also carry projections $\pm 4$ of the angular momentum 4. Consider a rotation by angle $\phi$ around the z-axis.
The $\sigma_+$ is transformed as $\sigma_+ \to \sigma_+e^{3i\phi}$, and
both ${\bm B}$ and ${\bm k}$ are usual vectors, i.e. 
$B_+\to B_+e^{i\phi}$, $k_+\to k_+e^{i\phi}$. The spherically symmetric Zeeman Hamiltonian (3) in the main text is invariant under the rotation (the U(1) symmetry).
However, the $\eta$-part of the effective Zeeman Hamiltonian must  transform as $e^{\pm 4i\phi}$. Since the $\eta$-part is not rotationally invariant, it depends on the choice of axes. Hereafter we assume that x and y axes are defined by crystal axes (100) and (010). The z-axis is (001). These arguments unambiguously fix the following most general kinematic form of the $\eta$-part of the  effective 2D Zeeman Hamiltonian
\begin{eqnarray}
\label{han1}
H_{\eta}& =& -\frac{\mu_B}{4}\left\{ {\overline {\delta_1}}k^2 \left( B_+\sigma_+ + B_-\sigma_-\right)  \right.\nonumber \\
&+&{\overline {\delta_2}} \left( k_-^2 B_+\sigma_- + k_+^2 B_- \sigma_+ \right)  \nonumber \\
&+& {\overline {\delta_3}} \left( k_+^6 B_+\sigma_- + k_-^6 B_- \sigma_+ \right) \nonumber  \\
&+&\left. {\overline {\delta_4}} \left( k_-^8 B_+\sigma_+ + k_+^8 B_-\sigma_-\right)\right\} \ .
\end{eqnarray}
The $\delta_1$ and $\delta_2$ terms have already been considered~\cite{komijani}.
The $\delta_3$ and $\delta_4$ terms in (\ref{han1}) are proportional to high powers of momentum and arise only in very high orders of the
perturbation theory. Therefore, they are small at values of k we are interested in, 
and we neglect these terms. The $\delta_1$ and $\delta_2$ terms arise in the second order
of perturbation theory. Repeating the calculations of Ref.~\cite{komijani} with 
the corrections pointed out below
we find the explicit second order perturbation theory expressions.
\begin{eqnarray}
&&{\overline {\delta_1}} = \eta {\overline Z_2}\nonumber\\
&&{\overline {\delta_2}} = \eta({\overline Z_1}-{\overline Z_3})\nonumber\\
&&{\overline Z_1}= \frac{3}{2}\kappa{\overline \gamma_2}Z_1\nonumber\\
&&{\overline Z_2}=6{\overline \gamma_2^2}Z_2\nonumber\\
&&{\overline Z_3}= \frac{3}{2}{\overline \gamma_2^2}Z_3 \ ,
\label{app1}
\end{eqnarray}
where $Z_{1,2,3}$ are the second order sums:
\begin{eqnarray}
\label{ZZ1}
&&Z_1 = -\frac{2}{m} \sum\limits_{n = 1}^{\infty} \frac{|\langle H1 | Ln \rangle|^2}
{\varepsilon_{Ln} - \varepsilon_{H1}}, \nonumber \\
&&Z_2 = \frac{2i}{m} \sum\limits_{n = 1}^{\infty} \frac{\langle H1 | (z-z_0) | Ln \rangle \langle Ln| p_z | H1 \rangle}
{\varepsilon_{Ln} - \varepsilon_{H1}},\nonumber\\
&& Z_3 = -2 i \sum\limits_{n = 1}^\infty \frac{\langle H1 | \{(z-z_0), p_z\} |Ln \rangle 
\langle Ln | H1 \rangle}{m(\varepsilon_{Ln} - \varepsilon_{H1})} \ .
\end{eqnarray}
Here $\varepsilon_{H1}$ is energy of the ground heavy hole state and $\varepsilon_{Ln}$ is 
energy of the $n^{\textrm th}$ light hole state, and both states are taken at $k = 0$; 
$|H1 \rangle$ and $|Ln \rangle$ are the corresponding
wave functions. 
While the idea of this calculation repeats Ref.~\cite{komijani}, our result is significantly
different from that of~\cite{komijani}.
There are two reasons for the difference. 
(i) The  $Z_3$ term is missing in \cite{komijani}. $Z_3$ is zero in an infinite 
rectangular quantum well, but it is nonzero for other shapes of the quantum well, see discussion in Ref.~\cite{Miserev1}.
In particular, $Z_3$ is important when an electric field $E_z$ is imposed.
(ii) Another point that has to be corrected compared to \cite{komijani} is related to the gauge invariance discussed 
in the main text of our paper.
The sums $Z_2$ and $Z_3$ in Eq.(\ref{ZZ1}) contain matrix elements of $(z-z_0)$. Hence, they are sensitive to
the choice of $z_0$. Varying $z_0$ one can get arbitrary values for these sums.
Therefore, it is absolutely important to use the  value of $z_0$ that is dictated by the gauge condition (8) in the main text. 
Disregard of this condition leads to wrong values of the subleading terms discussed in the present
appendix, and, of course, it would also lead to wrong values of the leading terms discussed in the
main text.

\begin{figure}[h]
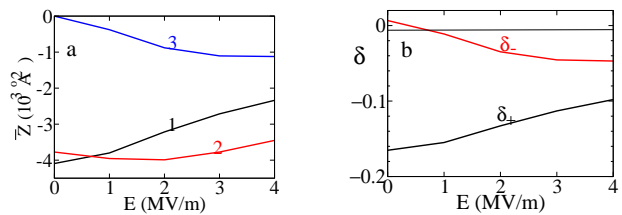

	\includegraphics[width=0.2\textwidth]{ZZ.eps}
\hspace{20pt}
	\includegraphics[width=0.2\textwidth]{dpm.eps}
	\caption{Panel a: Coefficients ${\overline Z}_1$, ${\overline Z}_2$, and ${\overline Z}_3$
versus electric field $E$.
Panel b: The $\eta$-corrections $\delta_{\pm}$ versus electric field $E$, the momentum is
$k=0.83 \times 10^{-2}$\AA$^{-1}$.
Both panels correspond to  infinite rectangular well of width 15nm in GaAs. 
The superimposed electric field is presented in units $10^6$V/m.
	}
	\label{ZZ}
\end{figure}

The coefficients ${\overline Z}_i$ have dimension of length squared, these coefficients  calculated
for infinite rectangular well of width $d=15$nm in GaAs are plotted in Fig.\ref{ZZ}a as functions of applied electric field.
It is worth noting that the  $\delta_1$ term in Eq.(\ref{han1}) 
has been probed in the optical experiment~\cite{Marie} for a symmetric quantum well.
This measurement does not give
a value of ${\overline {\delta_1}}$, but it allows to impose an upper limit on the
value. The upper limit is equal to our theoretical prediction that follows from Fig.\ref{ZZ}a.
In this sense, the theory is consistent with Ref.~\cite{Marie}.

For the QPC analysis it is convenient to introduce the following combinations.
\begin{eqnarray}
\label{ds}
\delta_+=k^2 ({\overline {\delta_1}}+{\overline {\delta_2}}) \nonumber\\
\delta_-=k^2 ({\overline {\delta_1}}-{\overline {\delta_2}})
\end{eqnarray}
Values of $\delta_{\pm}$ at $d=15$nm and $k=0.83 \times 10^{-2}$\AA$^{-1}$ are plotted in Fig.\ref{ZZ}b
versus applied electric field.

Assume that the x-axis (the QPC axis) makes angle $\phi$ with crystal axes,
\begin{equation}
{\hat x}=(\cos\phi,\sin\phi,0) \ .
\end{equation}
 Then using the same logic as that in the main text
and also using Eq.(3) and Eq.(\ref{han1}) we find the following expressions for 
the QPC g-factors
\begin{eqnarray}
\label{qpg}
&&|g_\parallel(\phi)| = \sqrt{\left(g_{\parallel}^{(0)} + \delta_- \cos 4\phi\right)^2 + \delta_-^2\sin^2 4\phi} 
\nonumber \\
&&|g_\perp(\phi)| = \sqrt{\left(g_{\perp} - \delta_+ \cos 4\phi\right)^2 + \delta_+^2\sin^2 4\phi},
\end{eqnarray}
Neglecting the $\delta_\pm^2$ terms that are proportional to $\eta^2$ and hence are beyond accuracy
of the theory, we can write
\begin{eqnarray}
\label{qpg1}
&&g_\parallel(\phi) =g_\parallel^{(0)}+  \delta_- \cos 4\phi \nonumber \\
&&g_\perp(\phi) =g_\perp^{(0)}- \delta_+ \cos 4\phi \ .
\end{eqnarray}

\renewcommand{\theequation}{C\arabic{equation}}
\renewcommand{\thefigure}{C\arabic{figure}}
\setcounter{equation}{0}
\setcounter{figure}{0}

\section{C: 3D distribution of the electrostatic potential and field in the device}
Here we present a full solution of the 3D electrostatic problem 
using the Thomas-Fermi-Poisson method, see for 
example~\cite{pyshkin,davies,modeling}. 
We solve numerically the electrostatic Poisson equation

\begin{equation}
\label{PE}
\nabla [\epsilon({\bm r})\nabla \varphi({\bm r})]=-\rho({\bm 
r})/\epsilon_0.
\end{equation}
Here 
$\varphi({\bm r})$  is the electrostatic potential,  $\epsilon({\bm r})\epsilon_0$ is 
the dielectric constant, and $\rho$ is the charge density.
There are three values of the dielectric constant, for GaAs $\epsilon=13$,
for AlGaAs $\epsilon=12.1$, and for Al$_2$O$_3$  $\epsilon=8$.
There are two contributions to the charge density $\rho$.
The first one comes from the 2DHS  $\rho_1=|e|n({\bm r})$, where  $e$ is the elementary charge, 
and  $n({\bm r})$ is the 
number density of holes in 2DHS. 
The number density $n({\bm r})$ is determined in 
Thomas-Fermi approximation in terms of the local 2D Fermi wave 
number $k_F$ 
\begin{eqnarray}
\label{TF}
n({\bm r})=|\Psi(z)|^2n_{2D}(x,y)=|\Psi(z)|^2\frac{k_F^2(x,y)}
{2\pi} \ ,
\end{eqnarray}
where $\Psi(z)$ is the wave function of the ground state  in 
the quantum well. 
The potential energy of electron is $U^{(e)}=-|e|\varphi$ and the 
potential energy of hole is $U=-U^{(e)}=|e|\varphi$.
To fix the reference level we choose $U=0$ in the 2DHS far away from the QPC.
Hence, the value of $k_F$ is determined by the equation
\begin{eqnarray}
\label{TF1}
\varepsilon_k-U({x,y})=\varepsilon_F \ ,
\end{eqnarray}
where $\varepsilon_k$ is the hole dispersion and 
$\varepsilon_F$ is the Fermi energy of the 2DHS.
Eq.(\ref{TF1}) makes sense only if it has solution with $k\geq 
0$, otherwise $n({\bm r})=0$.
Another contribution to the charge density comes from the interface
between GaAs and Al$_2$O$_3$, $\rho_2=|e|N_0\delta(z-z_1)$ where
\begin{eqnarray}
\label{n0}
N_0=5.5 \times 10^{11}cm^{-2} 
\end{eqnarray}
In the calculation we use
the real geometry of the structure. The geometry and the sizes 
are presented in Section A of the supplementary 
material.  Eq.(\ref{PE}) is solved with boundary 
conditions determined by potentials at the gates, the top gate, 
the back gate and the split-gate 
\begin{eqnarray}
\label{Ugate}
&&U_{tg}^{(e)}=-|e|V_{tg}-\Delta\nonumber \\  
&&U_{sg}^{(e)}=-|e|V_{sg}-\Delta,\nonumber \\
&&U_{bg}^{(e)}=-|e|V_{bg}-E_g,  
\end{eqnarray}
where $E_g=1.5$~eV is the band gap in GaAs, and 
\begin{equation}
\label{d}
\Delta\approx 0.6 eV 
\end{equation}
is a parameter describing the difference in values of workfunctions of the gate metal
and semiconductor. The value of $\Delta$ was fitted to describe the experimental
dependence of 2DHG density on $V_{tg}$ and $V_{bg}$.
We have to say that the set of parameters (\ref{n0}), (\ref{d}) can be changed,
for example one can take $N_0=0$ and $\Delta=0.85$eV. This also reproduces
the experimental dependence of  2DHG density on $V_{tg}$ and $V_{bg}$ but it results in some 
offset in  $V_{sg}$ compared to the experiment. In principle, the offset means nothing 
and we can work with the second set of parameters too, it does not influence our conclusions.
Nevertheless, we prefer to use the set (\ref{n0}), (\ref{d}) since it gives values
of $V_{sg}$ for  opening of particular 1D channels that are
reasonably close to experimental values, see Fig.1a,b in the main text.

The 2D hole dispersion $\varepsilon_k$ in Eq.(\ref{TF1}) is nonquadratic.
It is not convenient to work with such a dispersion in the electrostatic problem.
Therefore we solve the problem for two different quadratic dispersions with different masses,
\begin{eqnarray}
\label{III}
&& (i) \ \ \ \varepsilon_k=\frac{k^2}{2m^*}, \ \ \ m^*=0.13m_e\nonumber\\
&&(ii) \ \ \ \varepsilon_k=\frac{k^2}{2m^*}, \ \ \ m^*=0.3m_e \ ,
\end{eqnarray}
and check that practically there is no sensitivity to the dispersion.
In the first case the Fermi energy corresponding to the density $1.1 \times 10^{11}cm^{-2}$ is
$\varepsilon_F=2$meV, and in the second case $\varepsilon_F=0.87$meV.
Remarkably, the self-consistent potentials $U(x,y)$ in  these two cases are practically identical
after appropriate rescaling. So, the specific dispersion is not important. This is quite 
natural, since the potential is determined mainly by electrostatics, and quantum mechanics is 
of secondary importance.
Opening of the n$^{th}$ transverse channel is determined by Eq.(10) in the main text
where $E_n$ is eigenenergy of the Hamiltonian $k^2/2m^*+U(y)$.

\begin{figure}
\includegraphics[width=0.6\linewidth]{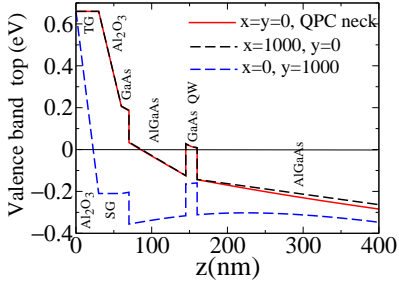}
\caption{Plots of tops of the valence band
versus z for setup S1 in the case when n=5  QPC transverse channel has
just opened (approximately at $V_{sg}=-0.4$V), the effective hole mass 
is $m^*=0.13$. 
The red line corresponds
to the neck of the QPC ($x=y=0$), the black dashed  line corresponds to a point deep in 2DHG 
($x=\infty$, $y=0$), and the blue dashed line corresponds to a point under the split-gate
($x=0$, $y=\infty$).
} 
\label{FS1}
\end{figure}

We have two experiments with different $E_z$ and gate biases described in Section A of the supplementary material, the parameters for setup 
S1 and setup S2 are summarized in the Table in Section A.
The result of selfconsistent solution of Eqs.(\ref{PE}),(\ref{TF}),(\ref{TF1}),(\ref{n0}),
(\ref{Ugate}),(\ref{d}),(\ref{III}) for setup S1 is plotted in Fig.\ref{FS1}.
Here we plot energy of the top of the valence band
 versus $z$ in the case when the n=5  QPC transverse channel has
just opened (approximately this corresponds to $V_{sg}=-0.4$V), the effective hole mass 
is $m^*=0.13$. 
The red line corresponds
to the neck of the QPC ($x=y=0$), the green line corresponds to a point deep in 
the 2DHS 
(say $x=\infty$, $y=0$), and the blue line corresponds to a point under the split-gate
(say $x=0$, $y=\infty$).

\begin{figure}
\includegraphics[width=0.49\linewidth]{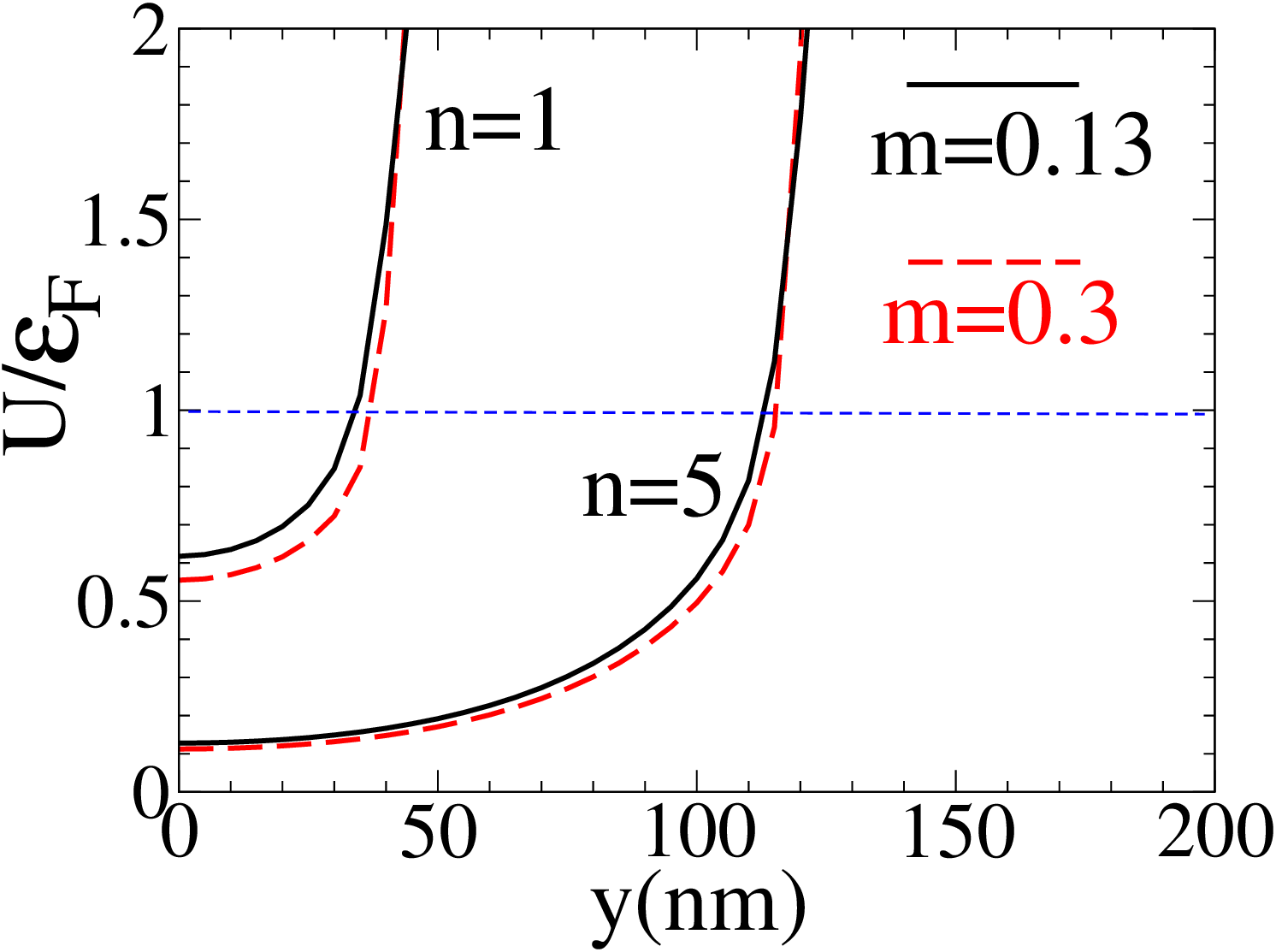}
\includegraphics[width=0.49\linewidth]{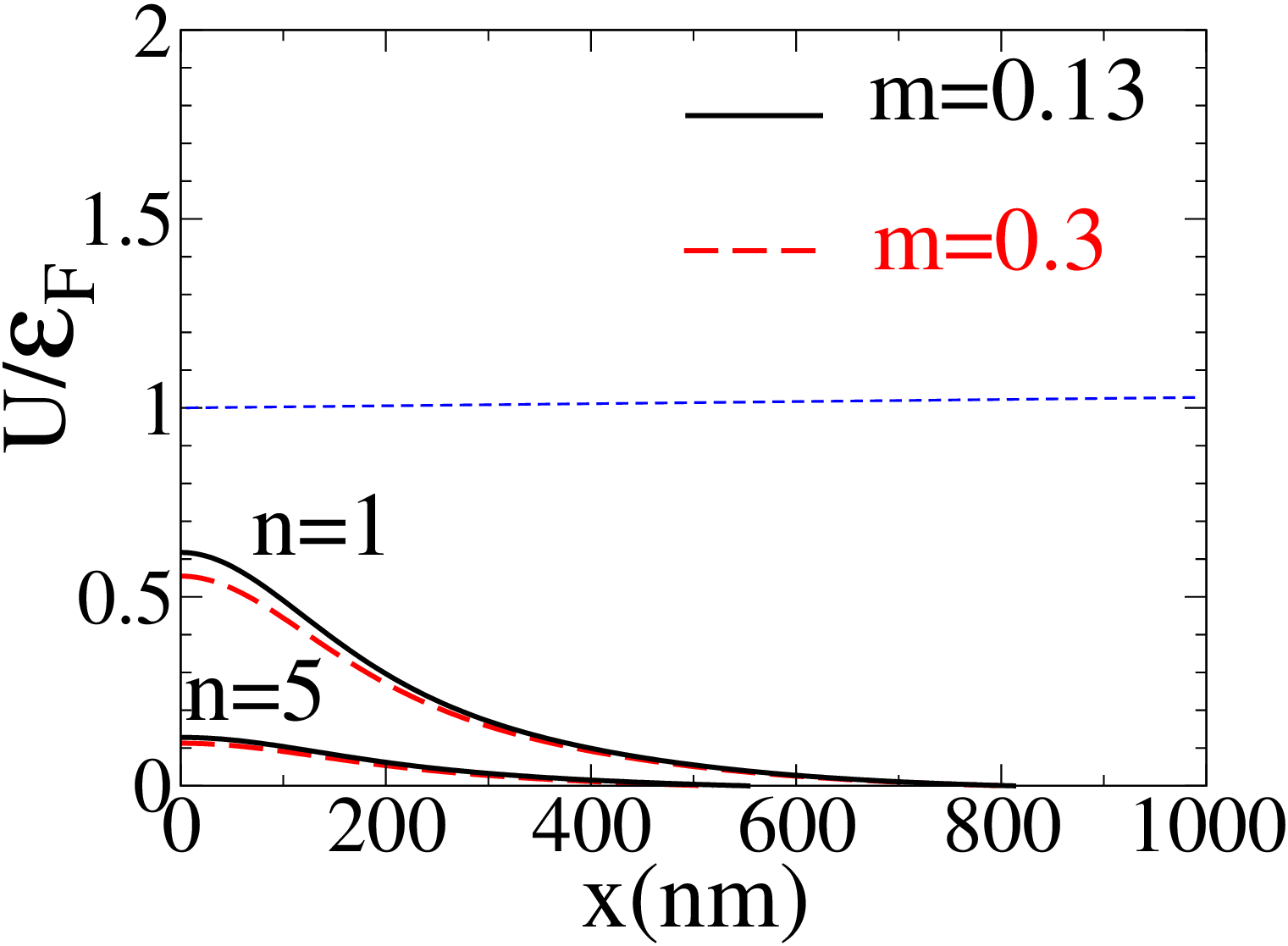}
\caption{Plots of $U/\varepsilon_F$ across (left panel) and along (right panel ) the QPC. 
Plots are presented for two effective masses and for two different transverse channels,
n=1 and  n=5 channels.
The picture is for the S1 setup. 
} 
\label{FSxyU} 
\end{figure}

Fig.\ref{FS1} gives the overall picture at a big scale of about 1eV.
Now we look at the scale about meV near zero.
We consider two dispersions presented in Eq.(\ref{III}).
The Fermi energy is 2meV in the case (i) and  0.87meV in case (ii).
The Fermi energy is the only low energy scale. Therefore in Fig.\ref{FSxyU}
we plot $U/\varepsilon_F$ across and along the QPC. The figure corresponds to the setup S1,
the figure for S2 is almost identical  to that for S1.
The curves are presented for the case when the n=1 channel has just opened ($V_{sg}\approx 0.1$eV) and for the case when the n=5 channel has just opened 
($V_{sg}\approx -0.4$eV).
This analysis justifies Fig.3a in the main text. 
Plots $U(y)$ presented in Fig3a correspond to $\varepsilon_F=2$meV 
since this is the Fermi energy corresponding to the real hole dispersion determined 
by Eq.(2) in the main text.

\begin{figure}
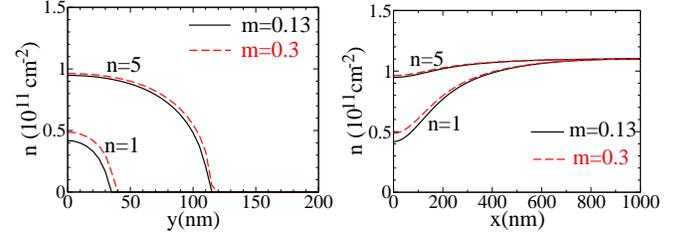

\includegraphics[width=0.49\linewidth]{FigyN.eps}
\includegraphics[width=0.49\linewidth]{FigxN.eps}
\caption{Plots of hole number density across (left panel) and along (right panel ) 
the QPC. 
Plots are presented for two effective masses and for two different transverse channels,
n=1 and  n=5 channels.
The picture is for the S1 setup. 
} 
\label{FSxyN} 
\end{figure}

In  Fig.\ref{FSxyN}
we plot the hole number density across and along the QPC. The figure corresponds to 
the setup S1, again the figure for S2 is almost identical  to that for S1.
The hole density at the QPC neck for high subbands is practically the same as that in 2D leads.

\begin{figure}
\includegraphics[width=0.49\linewidth]{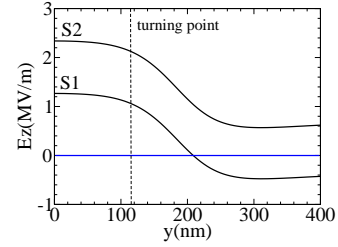}
\caption{The electric field $E_z$ across the QPC ($x=0$, $y \ne 0$) for setups S1 and S2.
These plots correspond to the case when the n=5 channel has just opened.
The plots for $m^*=0.13$ and $m^*=0.3$ coincide within the linewidth.
The vertical dashed line indicates the classical turning point determined from
Fig.\ref{FSxyU}left.
} 
\label{FSEz} 
\end{figure}

Finally, in  Fig.\ref{FSEz} we
plot the electric field $E_z$ across the QPC ($x=0$, $y \ne 0$) for setups S1 and S2.
The electric field at the neck of the QPC $x=y=0$ is practically equal to the field at 
$x=\infty$, $y=0$ (far away from the split-gates) presented in the Table in Section A of the supplementary material. 
We use this electric field for calculations of functions $g_1(k)$, $g_2(k)$,
$\delta_+(k)$, $\delta_-(k)$.

\renewcommand{\theequation}{D\arabic{equation}}
\renewcommand{\thefigure}{D\arabic{figure}}
\setcounter{equation}{0}
\setcounter{figure}{0}

\section{D:  Suppression of the transverse in-plane g-factor by combined action
of the QPC lateral confinement and the Rashba interaction}

The QPC is modeled here as a wire along the x direction, the momentum $k_x$ along the wire is zero.
The effective Hamiltonian describing the transverse channels then reads
\begin{equation}
\label{heR}
H\to \varepsilon_k-\alpha_kk^3\sigma_x +U(y)\ , \ \ \ k=k_y \ ,
\end{equation}
where $\varepsilon_k$ and $\alpha_k$ are even functions of $k$.
Assume for simplicity that $U$ is a rectangular well of width $L$.
The borders of the well are relatively smooth to make the $k^3$ term 
in Eq.(\ref{heR}) meaningful~\cite{com}.
Eigenstates of (\ref{heR}) have definite values $\sigma_x$, $|\uparrow\rangle$,
$|\downarrow\rangle$. The semi-classical wave functions then read
\begin{gather}
\psi_{\uparrow} = \sqrt{\frac{v_+v_-}{L(v_++v_-)}}\left(\frac{e^{ik_+y}}{\sqrt{v_+}}-\frac{e^{-ik_-y}}{\sqrt{v_-}}\right) 
\times |\uparrow\rangle \hphantom{d(A3)} \notag
\\
\psi_{\downarrow}  = \sqrt{\frac{v_+v_-}{L(v_++v_-)}}\left(\frac{e^{-ik_+y}}{\sqrt{v_+}}-\frac{e^{ik_-y}}{\sqrt{v_-}}\right) 
\times |\downarrow\rangle \ , 
\label{psi}
\end{gather}
where $k_{\pm}=Q \times (1\pm \delta_R)$ are solutions of the equation $\varepsilon=\varepsilon_k
\mp\alpha_kk^3$, and 
$\varepsilon$ is determined by  the quantization condition $QL=\pi n$, where $n=1,2,3,...$ enumerates transverse
channels. The group velocities are $v_{\pm}=\frac{\partial}{\partial k}(\varepsilon_k\mp\alpha_kk^3)$.
The Zeeman interaction, Eq. (3) in the main text, is an operator, $g\to \hat{g}$, for example, $\hat{g}[e^{ik_+y}-e^{-ik_-y}]=g(k_+)e^{ik_+y}-g(k_-)e^{-ik_-y}$.
Values of the QPC g-factors are given by the non spin flip and the spin flip matrix elements of the Zeeman interaction
\begin{gather}
G_{||}=\langle\psi_{\uparrow}|\hat{g}_{\||}\sigma_x|\psi_{\uparrow}\rangle=
\frac{1}{2}\left[g_{||}(k_+)+g_{||}(k_-)\right]\approx g_{||}(Q)\notag \\
G_{\perp}=i\langle\psi_{\downarrow}|\hat{g}_{\perp}\sigma_y|\psi_{\uparrow}\rangle
\approx \frac{\sin \pi n \delta_R }{\pi n \delta_R }\ g_{\perp}(Q) \ .
\label{gxy1}
\end{gather}
So, due to the Rashba SOI, $g_{\perp}$ has some additional oscillating suppression important at large n. 
In contrast, $g_\parallel$ remains constant at big $n$.
According to our estimates the value $\delta$ for quantum wells does not exceed 0.05.
Hence, this effect, somewhat similar to that considered numerically in Ref.\cite{Gelabert,kolasinski},
is much less important than the $g_1-g_2$ effect considered in the main text.